\documentstyle[twocolumn,epsf,aps]{revtex}
\draft
\begin{document}

\title{Critical domain size in a driven diffusive system}
\author{Attila Szolnoki, Tibor Antal, and Gy\"orgy Szab\'o}
\address{Research Institute for Materials Science, H-1525 Budapest,
POB 49, Hungary}

\address{\em \today}
\address{
\centering{
\medskip \em
\begin{minipage}{15.4cm}
{}\qquad The homogeneous ordered state transforms into a polydomain state  via
a nucleation mechanism in two-dimensional lattice gas if the particle
jumps are biased by an external field $E$. A simple phenomenological model
is used to describe the time evolution of a circular interface separating
the ordered regions. It is shown that the area of a domain increases if
its radius exceeds a critical value proportional to $1/E$ which agrees
qualitatively with Monte Carlo simulations.
\pacs{\noindent PACS numbers: 05.50.+q, 05.70.Ln, 64.60.Cn}
\end{minipage}
}}
\maketitle

\narrowtext

Lattice-gas models under the influence of an external (electric or
gravitational) field are often studied as a simple manifestation of
non-equilibrium systems \cite{kls,zia}. 
In these models the particle jumps are biased by a driving field $E$
resulting in particle transport through the system if periodic boundary
conditions are imposed. Our attention will be focused on a half-filled,
square lattice-gas model supposing repulsive nearest-neighbor interaction.
In the absence of a driving field ($E=0$) the lattice gas undergoes a
sublattice ordering when the system is cooled below the N\'eel
temperature. Dividing the lattice points into two interpenetrating
sublattices the ordered phases ($A$ and $B$) are usually described
by the average sublattice occupations. At low temperatures one of
the ordered phases dominates and its counterpart forms only small
disjoint domains (islands).

In driven lattice gases a similar ordering process was predicted
in earlier papers \cite{leung,dickman}. Very recently, however,
it has been demonstrated that at low temperatures a self-organizing
polydomain structure of the $A$ and $B$ phases, rather than a homogeneous
(monodomain) phase, is characteristic of this driven system in its stationary 
state \cite{szsza,szszab}. In other words, neither of the two homogeneous states
($A$ or $B$) is stable in the thermodynamic limit;
the monodomain structure transforms into a polydomain structure via a
nucleation mechanism. In Monte Carlo simulations the visualization
of particle distributions shows clearly that in the homogeneous $A$ (or $B$)
phase small domains of the $B$ ($A$) phase are created by thermal
fluctuations. The size of these domains fluctuates: they can either
grow or shrink, but typically collapse within some time. The size
of some domains can accidentally exceed the critical domain size, in which 
case the growth of these domains becomes definite. As a consequence,
sufficiently large domains spread throughout the system and the monodomain
phase transforms into a polydomain one. In the present paper we determine
the critical domain size using a simple analytical description and
Monte Carlo simulations.

The decay of a metastable phase via the nucleation mechanism is well known
in the theory of first order phase transitions\cite{gunton}. In such cases the
critical domain size is determined by the balance of surface and bulk
energies. Namely, the free energy difference between the stable and
metastable phases provides a driving force for the growth of stable regions
and the surface tension causes the minor regions to shrink. The situation in the
driven lattice gas is different because here the $A$ and $B$ phases are
thermodinamically equivalent. Here, the driving force of growth is produced 
by the enhanced interfacial material transport, which causes a rearrangement
of charges (extra particles and holes) along the interfaces. As a consequence, the
domains are polarized and the corresponding electrostatic force drives
the domain walls along the field. It will be shown that this mechanism
results in a field-dependent critical domain size.

For this purpose we adopt a simple phenomenological model used previously
to describe the interfacial instability \cite{szsza}. Neglecting noise 
we will concentrate on the deterministic motion. In this
phenomenological model the interface thickness and particle transport
in the bulk phases are also neglected. For later convenience 
we generalize the previous description to allow the appearance of
closed interfaces in the two-dimensional system.
Thus, the interface ${\bf r}$ and the interfacial charge density $\rho$
are described by single valued functions of time $t$ and a parameter
along the interface.

The continuity equation for the interfacial charge density obeys the
following formula:
\begin{equation}
{\partial_t \rho}=-\sigma{\bf E}\partial^{2}_{ss}{\bf r}
 +\sigma D \partial^{2}_{ss} \rho \,\,\,\, ,
\label{eq:charge}
\end{equation}
where $\partial_t$  and $\partial^{2}_{ss}$ denote the partial and
the second partial derivatives with respect to $t$ and $s$. Here it is
assumed that both the interfacial conductivity $\sigma$ and the diffusivity
$D$ are independent of the orientation. In the above expression we
have omitted the contribution caused by the variation of interface length
because it yields only a negligible correction to the solution.

The time evolution of the interface can be given as
\begin{equation}
\partial_t {\bf r}= C \partial^{2}_{ss}{\bf r}
 +\nu \rho({\bf E}- \partial_{s}{\bf r} ({\bf E} \partial_{s}{\bf r}))\,\,\, ,
\label{eq:wall}
\end{equation}
where the first term denotes the effect of surface tension and
the second describes the movement induced by the driving field.
Here the mobility of the interface is characterized by a parameter $\nu$.
Notice that $C$ and $\nu$ are also independent of the interface orientation.
In fact, the model parameters ($\sigma, D, C, \nu$) involve
temperature-dependence while field-dependence vanishes in the
low-field limit.

Starting from any initial interface and charge density
at $t=0$ the above equations define the interface evolution. For simplicity
we choose a circular domain to study the variation of its initial
shape and area over a short period. In polar coordinate system comformable
to the circular domain, the initial charge distribution $\rho$ is expressed
as a function of $\phi$ angle. The polar axis is parallel to the field and 
its origin is the centre of the domain. The initial charge distribution is
chosen to be the stationary solution of Eq.~(\ref{eq:charge}) for the fixed
initial radius ($r = R$). This assumption is consistent when determining the 
critical domain size, because the domain size remains unchanged in this state
for a sufficiently long time to reach the stationary charge distribution.
In the presence of a vertical field, the initial charge distribution 
may be expressed as a function of angle $\phi$, namely
\begin{equation}
\rho(\phi)={E \over D} R^2 \cos(\phi)
\label{eq:rho(phi)}
\end{equation}
where $E$ denotes the absolute value of the vertical field.
Notice that the total charge accumulated along the interface is zero
and the domain is polarized, that is, charges are transported from the
bottom semicircle to the top one. 
Substituting the above expression into Eq.~(\ref{eq:wall}) the time
derivation of radius $r$ at $t=0$ yields the following form:
\begin{equation}
\partial_t r (t=0) = {\nu \over D} E^2  R \cos^2(\phi) - 
{C \over R} \,\,\, .
\label{eq:dr/dt}
\end{equation}
If $E = 0$ then the circle contracts isotropically as expected. In the
presence of the driving field the first term is responsible for the
anisotropic increase. More precisely, in sufficiently strong fields
the circular domain elongates along the field, while its transversal
size is decreased by the surface tension. This anisotropic evolution
of domains is clearly observable in Monte Carlo simulations.
Here it is worth mentioning that the decrease of the transversal size is
proportional to the curvature at $\phi=\pi /2$ which tends to zero. In the
late stage of this evolution the part of the interface parallel to
the field remain unchanged \cite{szsza}.

Due to the anisotropic growth we can not adopt the concept of
critical domain size $R_c$ directly from the classical theories. 
In the present case, the definition of $R_c$ is based on the time
variation of the initial domain area, 
\begin{equation}
\partial_t A = \int\limits_{0}^{2\pi} R \partial_t r d{\phi}  
\,\,\,\, .
\label{eq:dA}
\end{equation}
which is an increasing function of the domain radius $R$. 
This quantity becomes positive if the initial radius exceeds the critical
domain size given as
\begin{equation}
R_{c} = {1 \over E} \sqrt{{2CD \over \nu}}\,\,\, .
\label{eq:cb}
\end{equation}
As a result the critical domain size as well as the life time of the
homogeneous metastable state diverge when $E \to 0$.

It should be emphasized that $R_c$ is directly related to the result of linear
stability analysis \cite{szsza}. That is, the planar interface perpendicular
to the field is unstable when the wave length of the periodic perturbations
is larger than $\lambda_0=\sqrt{2}\pi R_c$. The typical strip width was
approximated by the inverse of the wave number which is
characterized by the largest amplification rate \cite{szsza}. However, the
simple comparison of $R_c$ and the predicted strip width is not
possible because the prefactors of $1/E$ consist of different combinations
of model parameters. A more adequate picture can be drawn by finding
the explicit relationship between the model parameters and the
lattice-gas description. We intend to perform these calculations in the
near future.

It is worth mentioning that the $1 / E$ dependence of the critical domain
size has already been predicted by some authors on the basis of equilibrium
free energy arguments \cite{zia,mon}. Unfortunately, this approach also
contains phenomenological parameters whose adequate relation to the
parameters of the lattice-gas model is not known yet.

To check the above prediction we have performed a series of Monte Carlo 
simulations (details are given in previous works \cite{kls,binder}).
In our simulations the standard Kawasaki dynamics \cite{kaw} was modified
by taking  into consideration the effect of a driving field on a square lattice
with periodic boundary conditions \cite{szsza}. In each run the initial
state was created from a completely ordered phase $A$ in which we generated
a circular domain of $B$ phase by shifting the inside particles along the field 
by one lattice site. By this means, we obtained a "polarized domain" 
whose surface has an extra charge distribution similar to those given 
by Eq.~(\ref{eq:rho(phi)}). After $10-40$ Monte Carlo steps per particle
we have determined the area of the circular domain. To suppress the
thermal fluctuations this procedure was repeated several thousand times
at a given field and radius and a temperature ($T=0.44$) below
the peak of specific heat. The radius of the initial domains was chosen
from 8 to 200 on an $L \times L$ lattice where $L$ was increased 
simultaneously with $R$ from 36 to 600. 

From an average of the data obtained for various radii we were able
determine the critical domain size as a function of $E$. 
The results of the Monte Carlo simulations are illustrated in Fig.~1 where
the size of squares indicates the statistical error. These data confirm
qualitatively the theoretically predicted $1 / E$ behavior indicated
by a solid line in the figure. 

If we reinterpret the early calculations \cite{leung,dickman} the ordered
metastable states exist only below a ``critical temperature'' decreasing
with $E$. In other words, the number of defects (or the presence of $B$
domains) in the metastable $A$ phase increases with $E$ at fixed
temperatures. To avoid the discrepancy caused by the appearance of extra
defects, the above simulations have been carried out for weak fields
ensuring a low defect density in the metastable state.
Thus, the investigated range of the electric field may be
increased by decreasing the temperature. Unfortunately, in this case
we need longer run-times to have the same accuracy, which
makes the simulations more time-consuming.

\begin{figure}
\centerline{\epsfxsize=8.0cm
                   \epsfbox{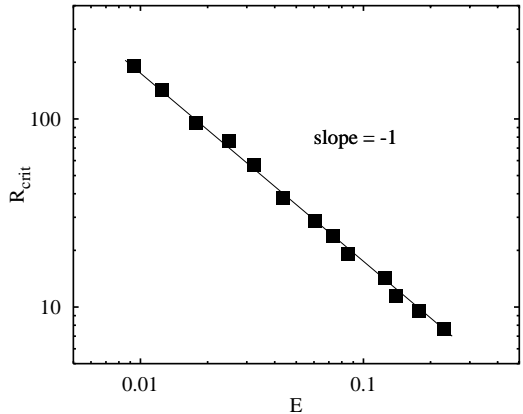}
                   \vspace*{2mm}      }
\caption{Critical domain radius {\it vs.} driving field at 
fixed temperature $T=0.44$.}
\label{fig:MC}
\end{figure}

In the above mentioned simulations we have chosen the radius to be
significantly larger than the interface thickness, wich is comparable with
the lattice constant. For smaller radii the deviation of the actual
interface from the circle becomes important and causes large fluctuations.
This phenomenon might have been a reason why the
verification of the $1/E$ behavior have not been successful in the early
Monte Carlo simulations \cite{teitel}.

In summary, we have explored the existence of a critical domain size
which is required for the homogeneous (metastable) ordered phase
to decay into a self-organizing polydomain structure via a 
nucleation mechanism. The phenomenological description of the interface
evolution suggests that the area of a circular domain increases if
its size exceeds a critical value. In qualitative agreement with
the Monte Carlo simulations the critical domain size varies inversely
with the field strength.

\vspace{5mm}

The authors thank Christof Scheele for careful reading of the manuscript.
This research was supported by the Hungarian National Research Fund
(OTKA) under Grant No. F-7240.

\end{document}